\documentclass[aps,superscriptaddress, onecolumn, longbibliography,12pt]{revtex4-1}

\usepackage[pdftex]{graphicx}

\usepackage{caption}
\usepackage[utf8]{inputenc}
\usepackage[T1]{fontenc}
\usepackage{mathptmx}
\usepackage{etoolbox}

\usepackage{amsmath}
\usepackage{amssymb}
\usepackage{ulem}

\makeatletter
\def\@email#1#2{%
 \endgroup
 \patchcmd{\titleblock@produce}
  {\frontmatter@RRAPformat}
  {\frontmatter@RRAPformat{\produce@RRAP{*#1\href{mailto:#2}{#2}}}\frontmatter@RRAPformat}
  {}{}
}%
\makeatother
\usepackage[utf8]{inputenc}

\begin{document}

\title{Universality in the tape-peeling trace}

\author{Keisuke Taga$^*$}
\email{tagaksk@akane.waseda.jp}
\affiliation{Department of Physics, Waseda University, Okubo 3-4-1, Shinjuku, Tokyo 169-8555, Japan}
\affiliation{Department of Systems and Control Engineering, Institute of Science Tokyo, Ookayama 2-12-1, Meguro, Tokyo 152-8550, Japan}

\author{Akihiko Toda}
\affiliation{Graduate School of Advanced Science and Engineering, Hiroshima University, Kagamiyama 1-4-1, Higashi-hiroshima, Hiroshima 739-8527, Japan}

\author{Yoshihiro Yamazaki}
\affiliation{Department of Physics, Waseda University, Okubo 3-4-1, Shinjuku, Tokyo 169-8555, Japan}

\date{\today}

\begin{abstract}
  \textbf{Spatiotemporal patterns, which are of interest in statistical physics and nonlinear dynamics, form on the tape-peeling trace.
Recently, we have proposed a mathematical model to describe these pattern formation in the tape-peeling trace.
In this paper, we further investigate the tape-peeling model from the perspective of its universality class.
We confirm that our model belongs to the 1-dimensional directed percolation universality class.
Furthermore, the experimental results from a previous study are re-analyzed, and it is suggested that the tape-peeling trace can also be classified within the 1-dimensional directed percolation universality class.}
  \end{abstract}
  \maketitle
\section*{Introduction}
 Adhesion is a ubiquitous phenomenon. One of the most familiar applications of it is the adhesive tape. 
 The utility of the adhesive tape lies in the ease of attaching and detaching objects.
 When detaching, we peel the tape. Although this process is familiar, closer observation reveals interesting physical phenomena that are too commonplace to be overlooked in the tape-peeling trace.

\begin{figure}[h]
  \centering
  \includegraphics[width=1\linewidth]{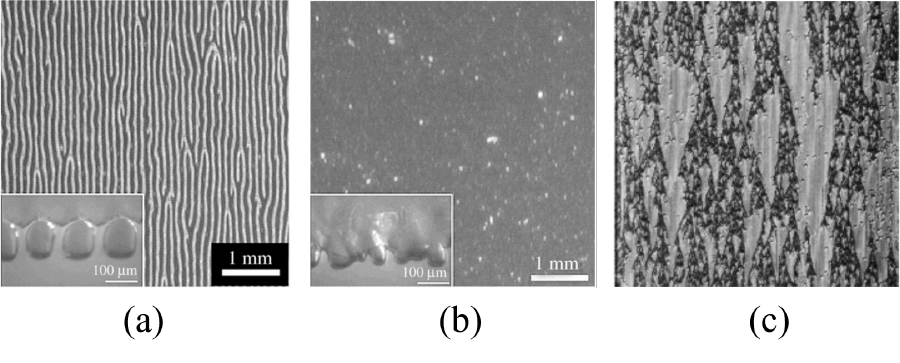}
  \caption{Pattern formation of the peeling trace~\cite{yamazaki2006patterns}. 
  (a) Slow peeling, (b) Fast peeling, (c) Intermediate speed peeling ($25\mbox{mm}\times 25\mbox{mm}$).}
  \label{f:experimental pattern}
\end{figure}

By paying attention onto the adhesive tape after peeling, various pattern formations may be found~\cite{Yamazaki2002-uw,sun2022rateDependent}.
Two of the authors have studied the properties of peeling dynamics and pattern formation of the peeling trace experimentally~\cite{Yamazaki2002-uw,Yamazaki2004-hw,yamazaki2006patterns,yamazaki2012spatiotemporal}.
It is found that the patterns of the peeling trace are related to the peel-front structures of the deformed adhesives, and these structures change by the peel speed~\cite{urahama1989peelload,yamazaki2006patterns}.
Figures~\ref{f:experimental pattern} show the typical peeling traces and related structures of the peel front.
The figure in the lower-left corner of  Fig.~\ref{f:experimental pattern}(a) corresponds to the tunnel structure for the slow peeling. In this peel-speed region, air penetrates the adhesives, forming a tunnel-like structure due to the fingering instability~\cite{ghatak2000instability}.
The tunnel persists in the adhesives after peeling, resulting in a stripe pattern in the trace, as shown in Fig.~\ref{f:experimental pattern}(a). The microscopic tunnel structure appears macroscopically white.
For the fast peeling, the tunnel structure collapses, as shown in the figure in the lower-left corner of Fig.~\ref{f:experimental pattern}(b), and the peeling pattern becomes black, as shown in Fig.~\ref{f:experimental pattern}(b).
And in the intermediate peel-speed region, these two peel-front structures switch, then black and white regions are mixed in the trace. Figure~\ref{f:experimental pattern}(c) shows one of the typical mixed patterns, which resembles a Sierpinski-gasket-like fractal random pattern. 
As the peel speed increases, the ratio of the black region grows, and eventually, the whole region is absorbed into the black.

Some models have been proposed and investigated to elucidate the mechanism of this pattern formation~\cite{Yamazaki2004-hw,yamazaki2006patterns,yamazaki2011bistable,yamazaki2012spatiotemporal,ohmori2019comments,taga2023peeling,taga2024peeling}. The following model is one of the peeling models that we proposed recently~\cite{taga2023peeling}:

\begin{align} 
  \frac{d^2 u_i}{d t^2}=-\frac{d U(u_i)}{d u_i}-b\left(V+\frac{d u_i}{d t}\right)+\frac{a}{1+c(u_i-d)^2}\frac{d u_i}{d t}\cr
  +D_1(u_{i+1}+u_{i-1}-2u_i)+D_2\left(\frac{d u_{i+1}}{d t}+\frac{d u_{i-1}}{d t}-2\frac{d u_{i}}{d t}\right),
\label{eq:general model}
\end{align}
where $u_i$ is the deformation of the $i$th unit at the peel front, $V$ is a peeling speed, $b$ is a viscous dissipation, $a$, $c$, $d$ are positive constants, and $D_1$ and $D_2$ are diffusion constants.

From the experiments, it is observed that the deformation of the peel front of the tunnel structure is larger than that of the without-tunnel structure~\cite{yamazaki2006patterns}, thus, we consider large $u_i$ as the tunnel structure and small $u_i$ as the without-tunnel structure. 
It can be considered that the peel front is discretized by the tunnel structures, so we discretize the spatial variable of Eq.~(\ref{eq:general model}) with the characteristic size of the tunnel structure.
To describe the bistability of the tunnel structure and the without-tunnel structure, we introduce the double-well potential $U(u)$ whose wells correspond to each structure.
The term $-b\left(V+\frac{d u_i}{d t}\right)$ relates to viscosity. 
$\frac{a}{1+c(u_i-d)^2}\frac{d u_i}{d t}$ is negative viscosity, introduced to represent slip-stick-like dynamics. 
$D_1$ and $D_2$ are introduced from the viscoelastic property of the adhesives.
Bubbles between the adhesives and a substrate, and inhomogeneity can break the tunnel structure as noise, so we also add stochasticity as $u_i\to 0$ with probability $p\ dt$ for this model.
In the following, we refer to this phenomenological model that describes the dynamics of the deformation of the adhesives at the peel front as the tape-peeling model.

For numerical simulation, we apply the following specific parameters to Eq.~(\ref{eq:general model})~\cite{taga2023peeling}:
\begin{equation}
  \begin{split}
  \frac{d^2u_i}{d t^2}=&-3(u_i-1)^2(u_i-2)-\left(V+\frac{d u_i}{dt}\right)+\frac{2}{1+20(u_i-1)^2}\frac{d u_i}{d t}\\
  &+(u_{i+1}+u_{i-1}-2u_i)+0.1\left(\frac{d u_{i+1}}{d t}+\frac{d u_{i-1}}{d t}-2\frac{d u_i}{d t}\right),\\
  &i=1,\ldots,N,
  \end{split}
  \label{eq:model}
\end{equation}
we used the $4$th-order Runge-Kutta method with a time step of $dt=0.01$ and applied periodic boundary conditions. 
The initial conditions are $u_i=0$ and $\frac{du_i}{dt}=0$. Additionally, noise is added as $u_i\to 0$ with a probability of $p\ dt=0.01 dt$~\cite{taga2023peeling}.

\begin{figure}[h]
  \centering
  \includegraphics[width=1\linewidth]{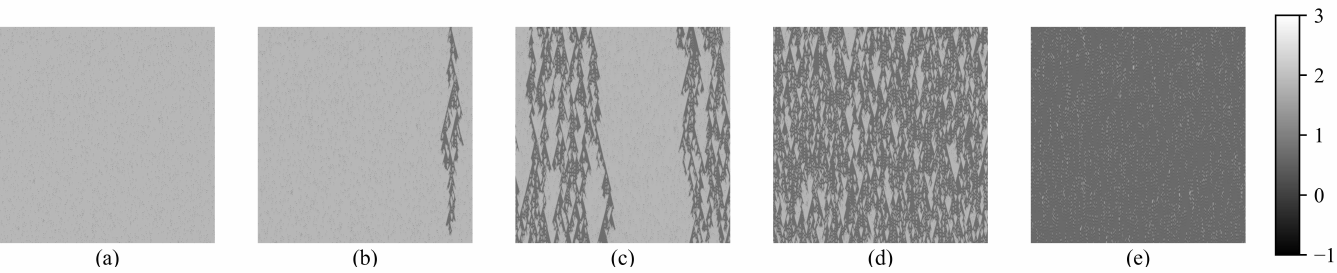}
  \caption{Patterns obtained from the numerical calculation of Eq.~(\ref{eq:model}). Time proceeds from top to bottom ($125 < Vt \leq 250$). The number of grids is $N=10^3$. (a) $V=0.28$, (b) $V=0.30$, (c) $V=0.309$, (d) $V=0.32$, (e) $V=0.50$.}
  \label{f:model pattern}
\end{figure}

By calculating Eq.~(\ref{eq:model}) with various $V$, we obtain patterns similar to the experimental patterns, as shown in Fig.~\ref{f:model pattern}.
As mentioned, the tunnel structure is related to the white region, and the without-tunnel structure is related to the black region, so we represent the large $u_i$ as white and small $u_i$ as black in the figures.
The qualitative appearance of the patterns is similar to the experimental results. Also quantitatively, some scaling properties of the pattern from the tape-peeling model are consistent with experimental results, indicating that the tape-peeling model can reproduce the pattern formation of the tape-peeling trace~\cite{taga2023peeling}. 
In this paper, we further investigate the tape-peeling model (Eq.~\ref{eq:model}) from the viewpoint of the universality class of spreading and diffusion phenomena. 

\section*{Derected percolation}
If we consider the white pattern (tunnel structure) and the black pattern (without tunnel structure) as states A and B, then the state of the peeling front transforms from state A to state B by changing the control parameter $V$. The transition can be considered a non-equilibrium phase transition.
The pattern formation in this phase transition evokes a relation to directed percolation (DP). DP is a universality class for non-equilibrium phase transitions of the spreading phenomena from a fluctuating state to an absorbing state~\cite{hinrechsen2009absorbing}. 

Systems in the DP universality class share critical exponents, allowing us to classify systems by measuring them. The number of independent exponents is 3. The conventional exponents are $\beta$, $\nu_\parallel$, and $\nu_\perp$, relating to the ratio of the absorbing state $\rho  \sim (V-V_c)^{\beta}$, the correlation time $\tau\sim (V-V_c)^{\nu_\parallel}$ and the correlation length $l\sim(V-V_c)^{\nu_\perp}$ where $V$ is a control parameter and $V_c$ is a critical point.
Many models are found to be related to DP, such as the Domany-Kinzel model~\cite{domany1981dp}, contact process~\cite{mollison1977contact}, and chemical catalysts~\cite{schlogl1972reaction}.
And, a few experimental realizations are reported for the turbulence transition of liquid crystals~\cite{takeuchi2007lc} and laminar flow ~\cite{lemoult2016couette,sano2016turbulence,klotz2022phase}.

It is conjectured that a system with the following properties shows the DP~\cite{janssen1981nonequilibrium,grassberger1982transition}.
\begin{enumerate}
  \item[I.] The system has a fluctuating phase and a unique absorbing phase.
  \item[II.] The transition point is a continuous phase transition.
  \item[III.] The transition can be characterized by a one-component order parameter.
  \item[IV.] The system has a short-range interaction.
\end{enumerate}
Considering tape-peeling phenomena, the system has a fluctuation from the bubbles in between adhesives and a base, and inhomogeneity of the adhesives. 
We can recognize the white state as a fluctuating state and the black state as a unique absorbing state (I). Also, considering the area of the black region $\rho$ as an order parameter, which increases gradually as the peeling speed $V$ increases, it shows a continuous phase transition (II,III)~\cite{yamazaki2006patterns}. 
Moreover, we can assume the interaction of the peel front as a short-range interaction for the stiff system (IV).
Therefore, peeling phenomenon may hold the above four properties. 
However, the exponents related to DP were not discussed in previous studies. Only $\rho$, which is related to the critical exponent $\beta$, is measured in the previous research but the relationship with DP was not discussed~\cite{yamazaki2006patterns}. 

In the following discussion, we report that the tape-peeling model (Eq.~\ref{eq:model}) has the property of the DP universality class. Additionally, we newly analyze those scaling exponents from the series of experiments Ref.~\cite{Yamazaki2002-uw,Yamazaki2004-hw,yamazaki2006patterns}. 
The results support that the tape-peeling trace may be regarded as an experimental realization of the DP universality class for 1D systems.

We numerically calculated Eq.~(\ref{eq:model}) and investigated the critical exponents related to the DP universality class. 
For the numerical calculation, we used the same setup for obtaining Fig.~\ref{f:model pattern}, with the system size $N=10^4$. We use the time-series data of $10^4<t\leq2\times 10^4$ for the analysis. To binarize the obtained patterns, we considered $u>1$ as the white region (state A) and $u\leq 1$ as the black region (state B). 
The DP universality class is characterized by some scaling exponents. The number of the independent exponents is 3; thus, we considered the following independent exponents $\beta$, $\mu^\parallel$, and $\mu^\perp$, which related to the density of the black region $\rho\sim (V-V_c)^{\beta}$, and the distribution of the gaps in the white region for the space $l$, $N(l)\sim l^{\mu^\parallel}$ and the time $\tau$, $N(\tau)\sim \tau^{\mu^\perp}$ at the critical point $V = V_c$~\cite{hinrechsen2009absorbing}.

\begin{figure}[h]
  \centering
  \includegraphics[width=1\linewidth]{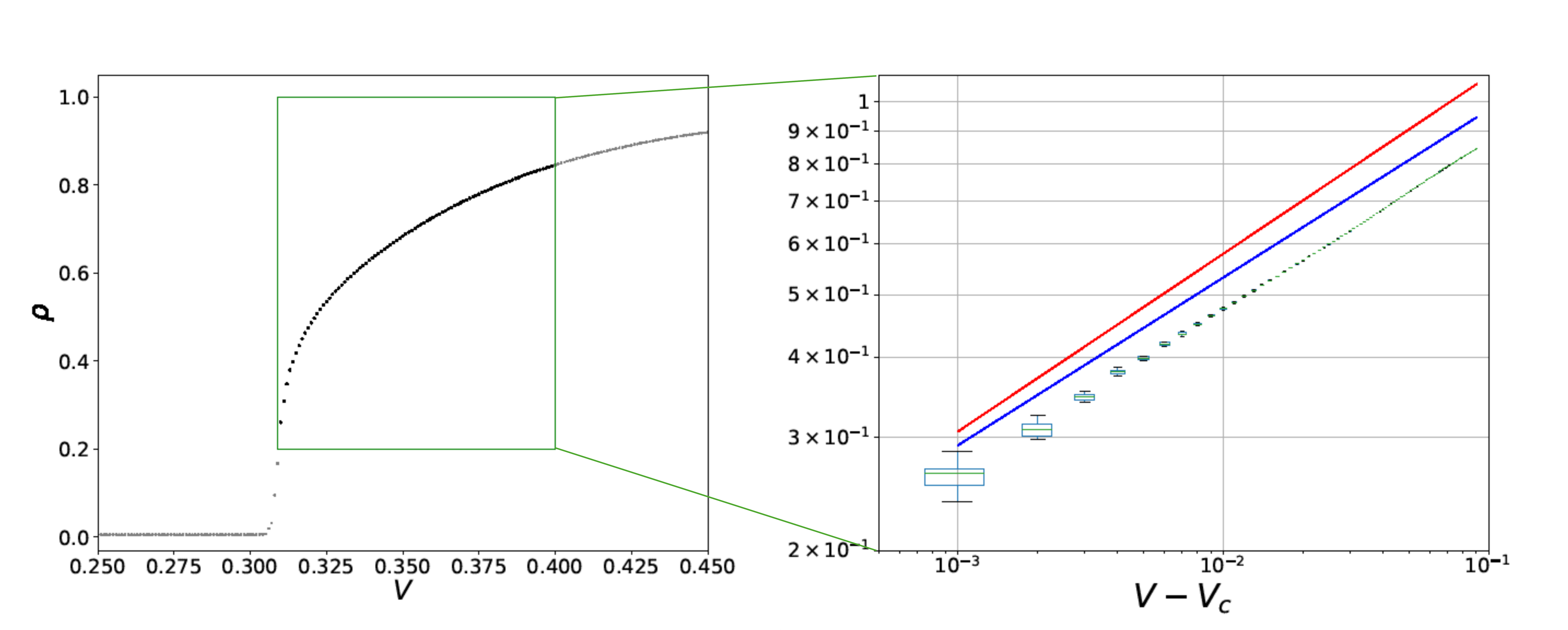}
  \\(a-1)\hspace{7cm}(a-2)\vspace{5mm}\\
\includegraphics[angle=90,width=0.47\linewidth]{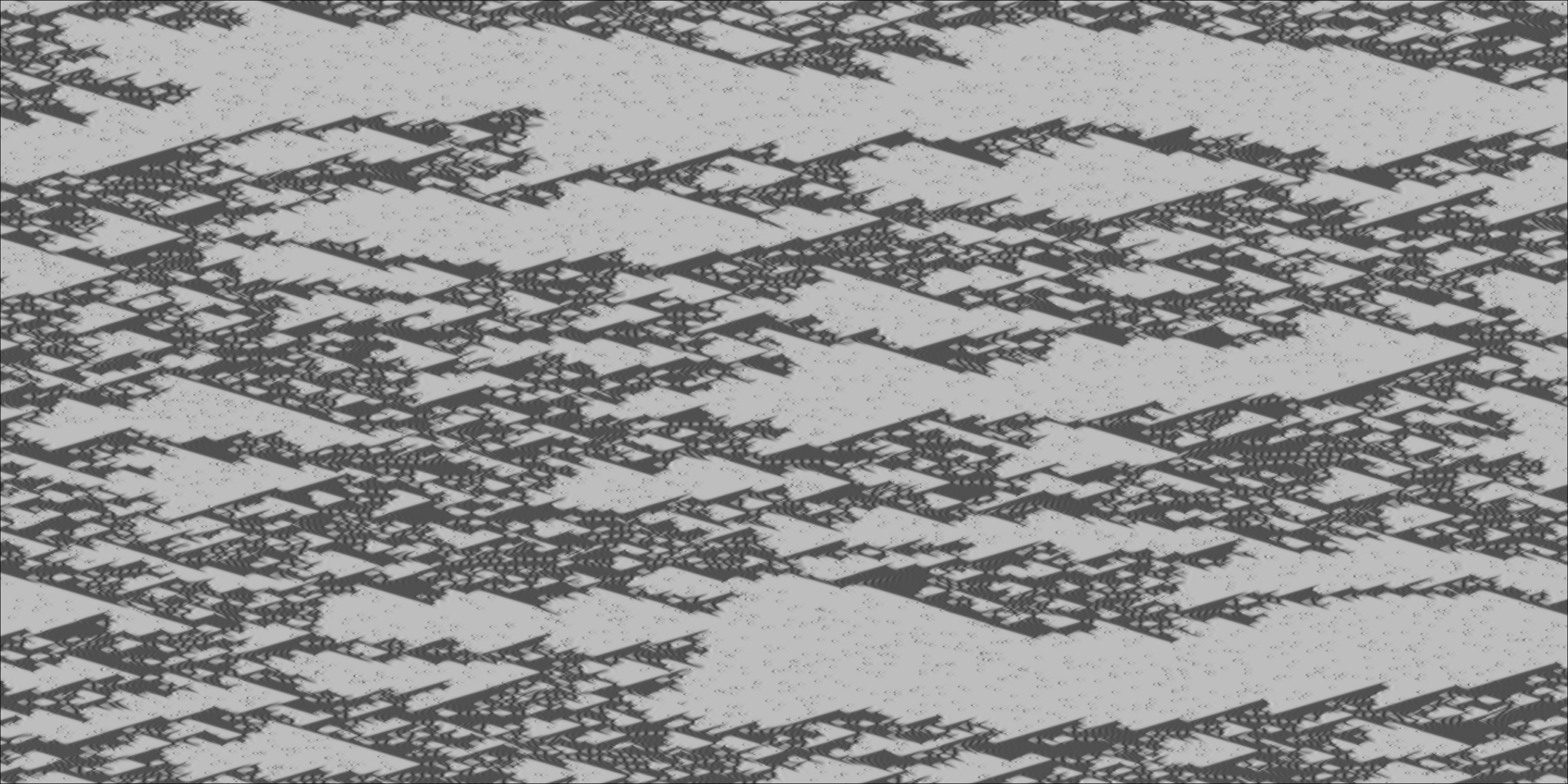}
  \includegraphics[angle=90,width=0.47\linewidth]{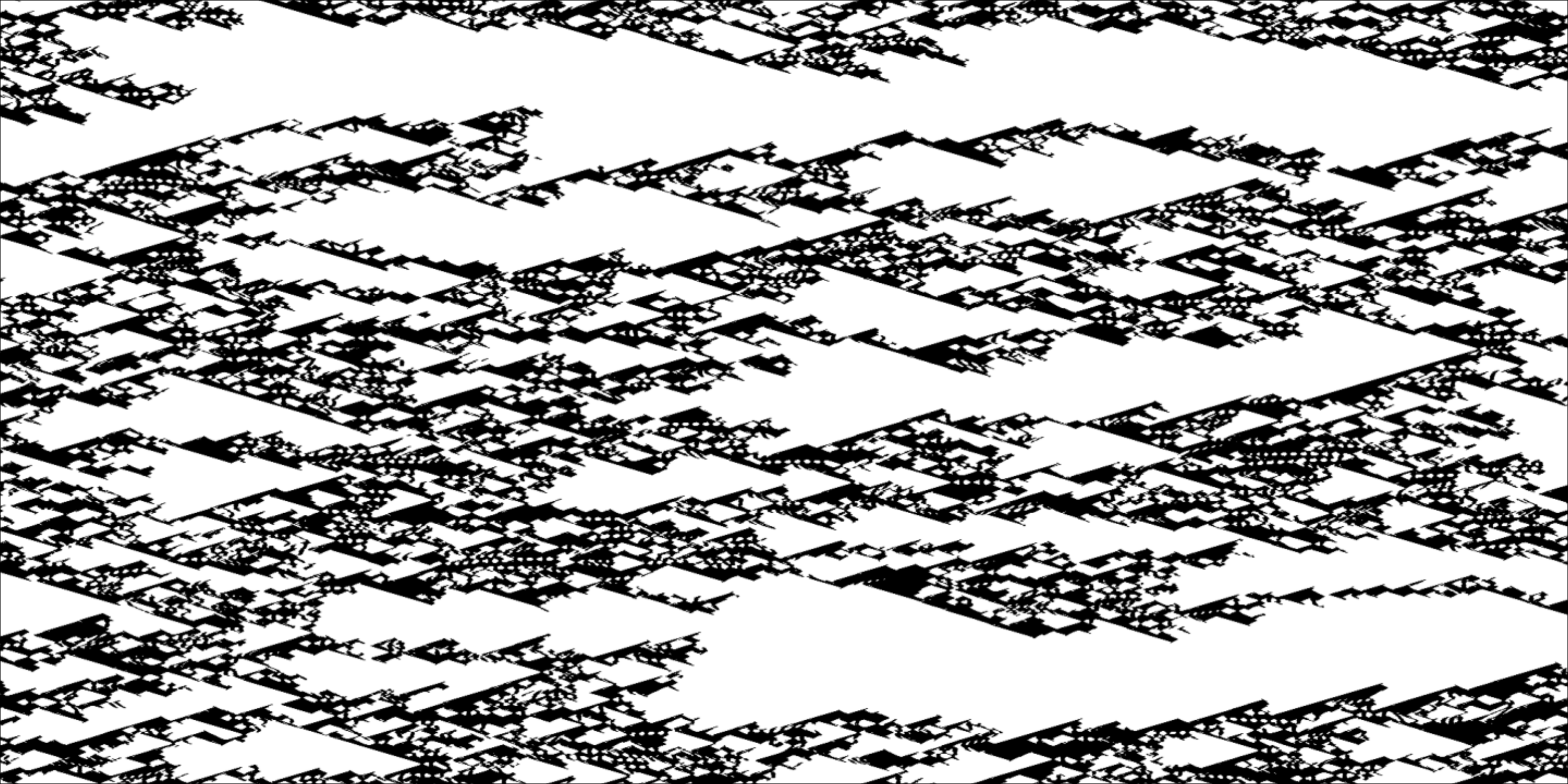}
  \\(b-1)\hspace{7cm}(b-2)\\
  \includegraphics[width=0.47\linewidth]{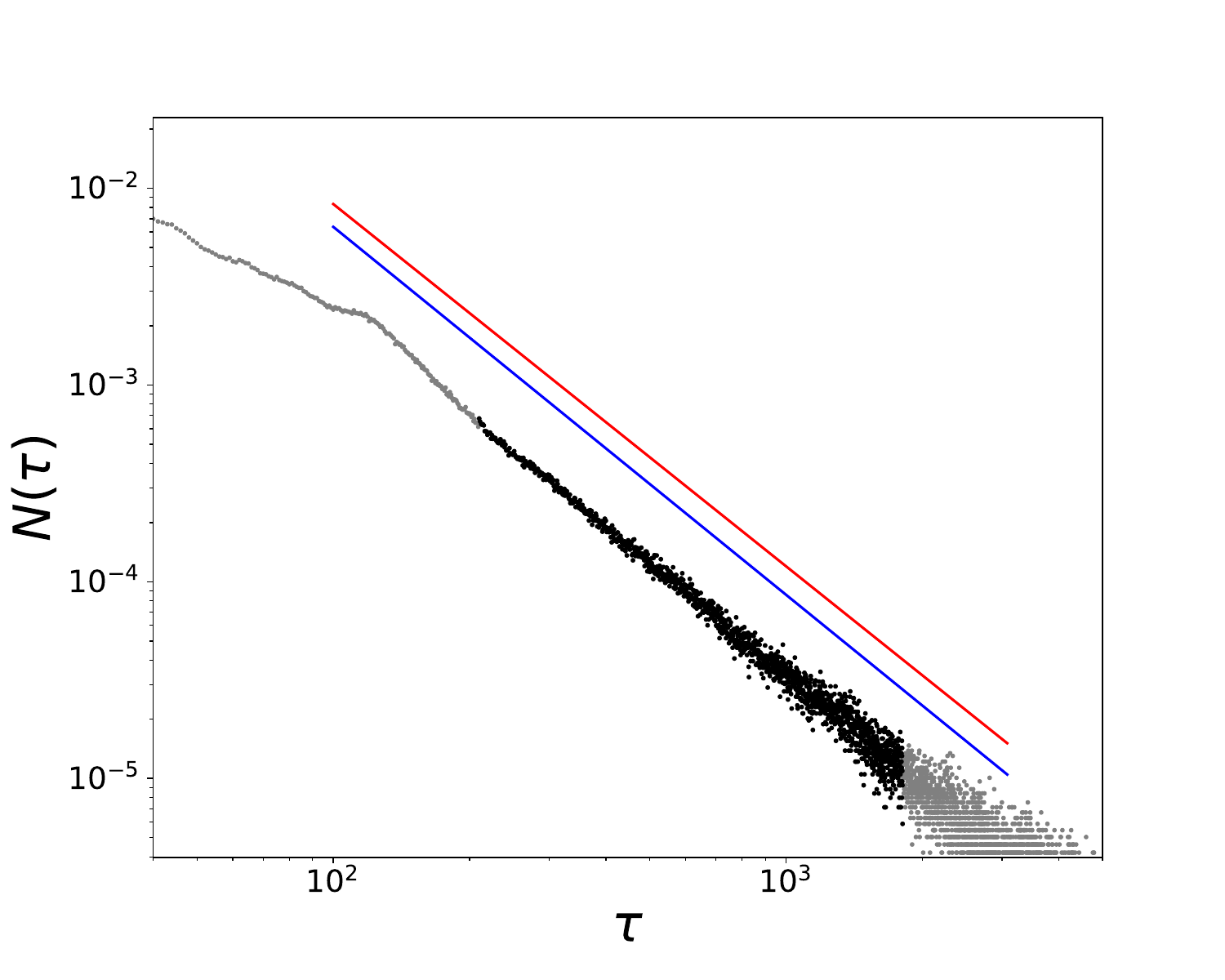}
  \includegraphics[width=0.47\linewidth]{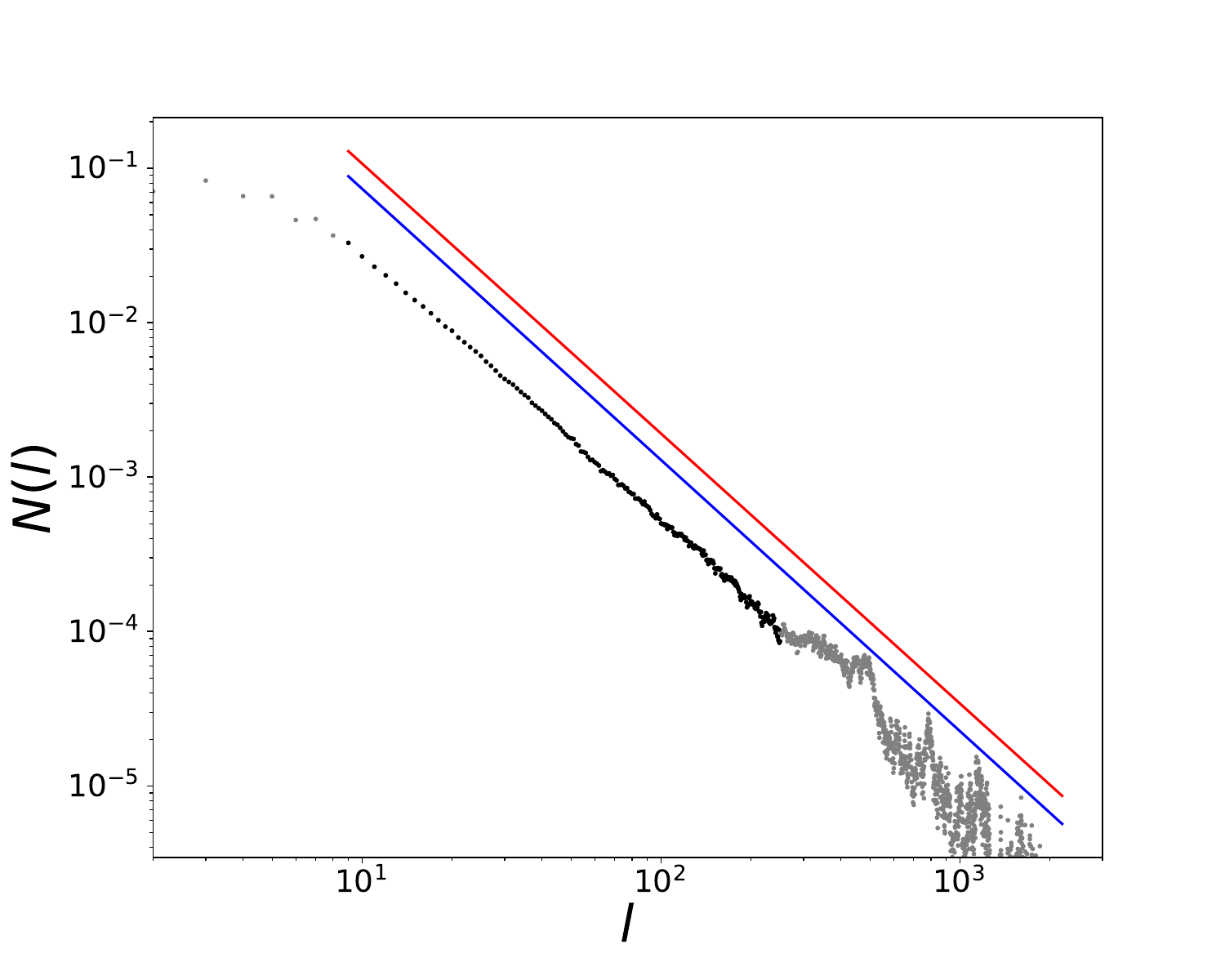}
  \\(c-1)\hspace{7cm}(c-2)
\caption{Scaling for the results from the tape-peeling model.\ (a) (a-1) The relationship between $\rho$ and $V$ obtained from the tape-peeling model with the average of 10 simulation runs. (a-2) Box plot on a log-log scale. The data are from 10 simulations. We set $V_c = 0.309$. Blue line represents the best fit of the average data, while red line serves as an eye guide with the slope of DP.\ (b) An example of the image processing. The figures illustrate binarization and denoising and were not used in the actual analysis. (b-1) A pattern with $V = V_c = 0.309$ obtained using the tape-peeling model with system size $N = 10^3$. Time proceeds from left to right ($10^3 < t \leq 2 \times 10^3$). (b-2) The binarized and denoised pattern of (b-1).\ (c) The distribution of (c-1) $N_\parallel$ and (c-2) $N_\perp$ for the tape-peeling model with $V=V_c = 0.309$. The scaling exponents are measured from the data shown with black dots. Blue lines represent the best fit, while red lines serve as eye guide with the slope of DP.}
  \label{fig:model}
\end{figure}

Figures~\ref{fig:model}~(a) show the relationship between $\rho$ and $V$. We fix the critical point as $V_c=0.309$ and obtained the critical exponent $\beta_{model}=0.262(16)$, which is similar to the theoretical value of DP; $\beta_{DP}=0.2765$~\cite{hinrechsen2009absorbing}.
In addition, we analyzed binarized and denoised patterns of the tape-peeling model (Fig.~\ref{fig:model}~(b)). We obtained the exponents $\mu^\parallel_{model}\sim 1.87(5)$ and $\mu^\perp_{model} \sim 1.76(3)$ as shown in Figs.~\ref{fig:model}~(c), which are consistent with DP values; $\mu^\parallel_{DP}\sim 1.841$ and $\mu^\perp_{DP} \sim 1.748$. These scaling exponents are summarized in Table I.
From the above results, it is confirmed that the tape-peeling model belongs to the DP universality class. Also, we may expect that the tape-peeling trace is an experimental realization of DP transition.

\section*{Discussion}
Then, our interest is whether the pattern formation of the tape-peeling trace in the real world falls within the DP universality class.
The details will be surveyed in future experiment; however, for now, we reanalyzed the experimental results which are obtained in the series of experiments for Ref.~\cite{Yamazaki2002-uw,Yamazaki2004-hw,yamazaki2006patterns} around 20 years ago. In the experiment, a PET film tape with $25\mbox{mm}$ width and $25\mu \mbox{m}$ thickness (No. 31D, Nitto Denko Corporation, Japan), which is now discontinued, was used.

\begin{figure}[h]
  \centering
\vspace{-5mm}
  \includegraphics[width=1\linewidth]{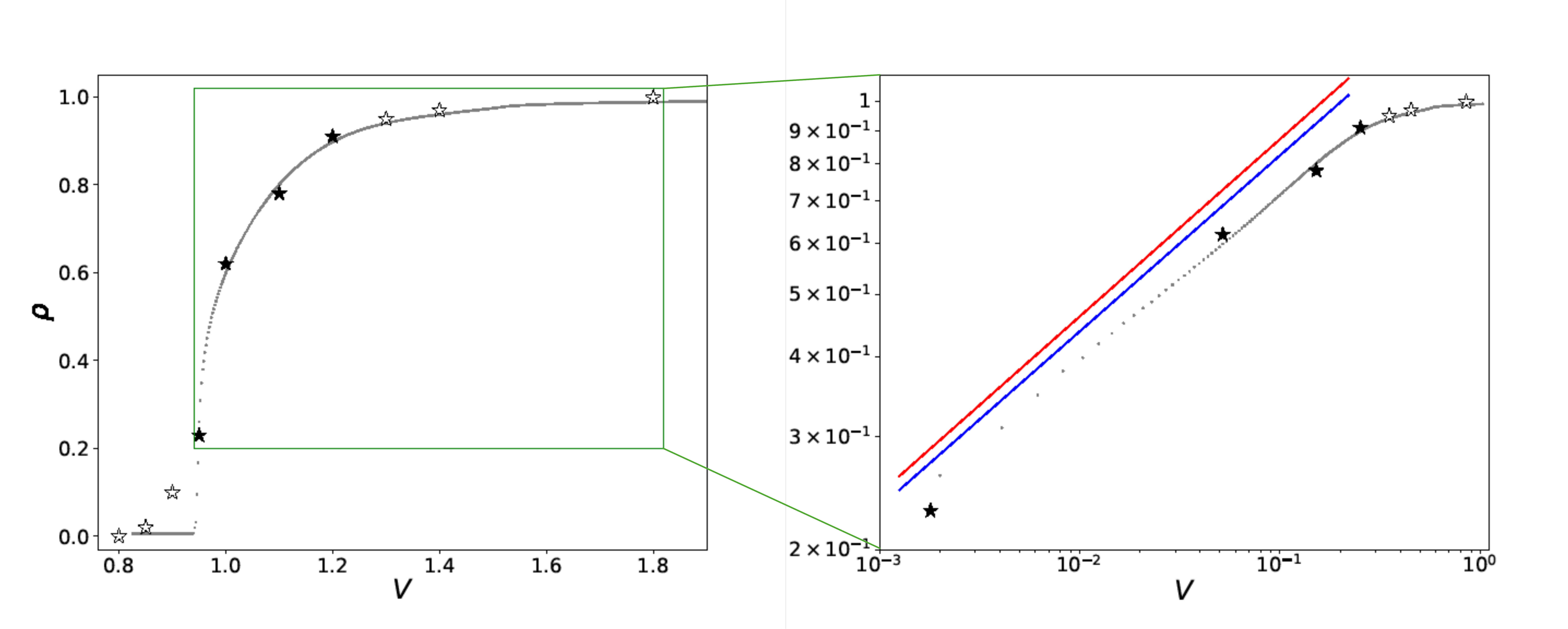}
  \\(a-1)\hspace{7cm}(a-2)\vspace{5mm}\\
    \includegraphics[width=0.9\linewidth]{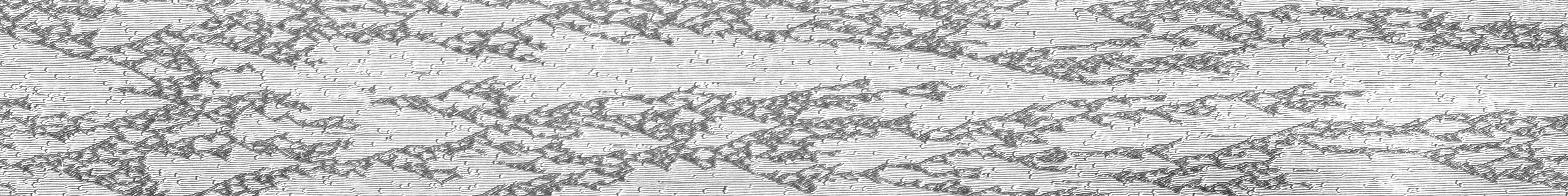}\\(b-1)\\\vspace{5mm}
  \includegraphics[width=0.9\linewidth]{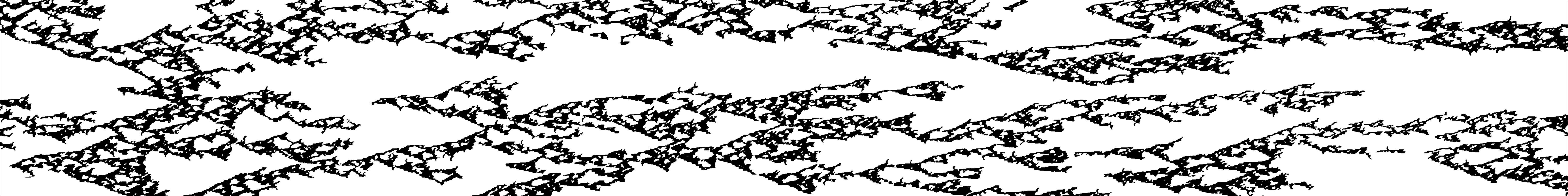}
  \\(b-2)\\
  \includegraphics[width=0.47\linewidth]{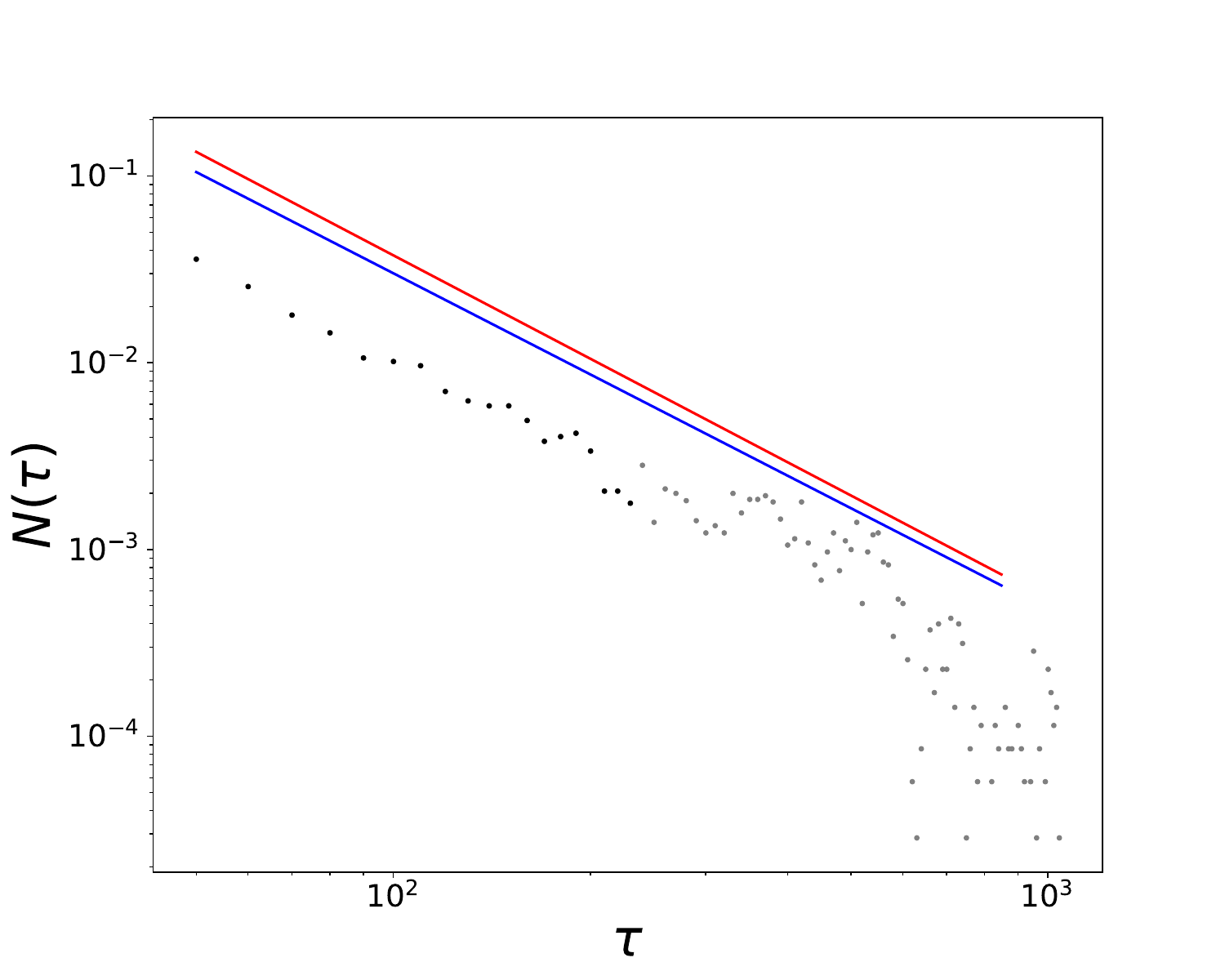}
  \includegraphics[width=0.47\linewidth]{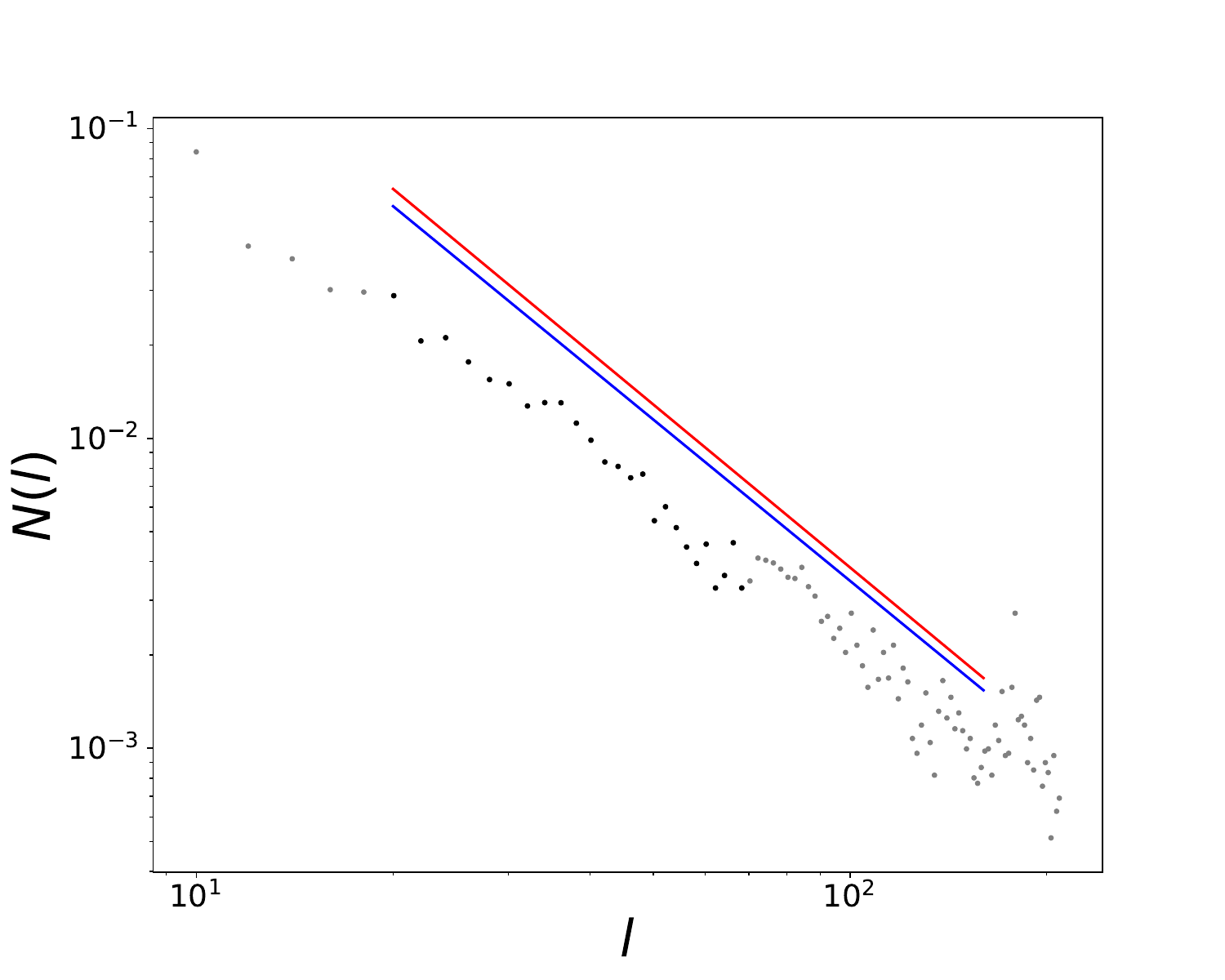}
  \\(c-1)\hspace{7cm}(c-2)
  \caption{Scaling for the results from experiments.\ (a) (a-1) The relationship between $\rho$ and $V$ obtained from the experimental result. (a-2) Log-log plot of (a-1). We set $V_c = 0.9483$. The scaling exponent is measured from the data shown with black stars. Blue lines represent the best fit and, red lines serves as an eye guide with the slope of DP. For the comparison, we plot averaged data shown in Fig.3 (a) with gray dots. The peeling speed $V^\prime$ for the model is translated and rescaled as $V = (V^\prime - 0.1461)/0.48$.\ (b) An example of the image processing. The figures illustrate binarization and denoising. (b-1) A pattern with $V = V_c \sim 0.46(\mbox{mm/min})$. Peeling proceeds from left to right. The width of the tape for this experiment was $25\mbox{mm}$, but because the shading of the scan data varied from place to place, so nearly uniform area was extracted. Thus, the real size of this figure is $12\mbox{mm}\times96\mbox{mm}$. (b-2) The binarized and denoised pattern of (b-1) for the analysis of (c).   
  \ (c) The distribution of (c-1) $N_\parallel$ and (c-2) $N_\perp$ obtained from (b-2). The scaling exponents are measured from the data shown with black dots. Blue lines represent the best fit and, red lines serves as an eye guide with the slope of DP.}
  \label{fig:exp}
\end{figure}

As shown in Fig.~14 of Ref.~\cite{yamazaki2006patterns}, the ratio of the white region $\hat{\rho}$ was measured for each $V$ at $25.5\pm0.5{}^\circ \mbox{C}$. $\hat{\rho}$ is related to the ratio of the black region $\rho$ as $\rho=1-\hat{\rho}$. 
We rescaled $\rho$ from the tape-peeling model and find that the tape-peeling model reproduces the experimental result as shown in Fig.~\ref{fig:exp}~(a), and the experimentally measured $\beta_{exp}= 0.275(29)$ is consistent with the value of DP. 

Physical values related to the following exponents, $\mu^\parallel$ and $\mu^\perp$, were not measured in the previous study. Thus, we newly analyzed the experimental patterns of the tape-peeling trace. We note that the experimental patterns used to obtain the ratio of white area were left, so another pattern measured at $17{}^\circ \mbox{C}$ was used. For this reason, a fractal pattern was obtained with the velocity $0.46 \mbox{mm/min}$, where the black region hardly appears in Fig.~\ref{fig:exp}~(a-1). 
From the analysis of the binarized and denoised experimental patterns (Figs.~\ref{fig:exp}~(b)), we obtained the exponents $\mu^\parallel_{exp} = 1.80(7)$ and $\mu^\perp_{exp} = 1.72(6)$ as shown in Figs.~\ref{fig:exp}~(c), which are also consistent with the DP values. 
These scaling exponents are also summarized in Table~\ref{tbl:exponents}.

\section*{Conclusion}
We investigated the tape-peeling model from the viewpoint of the universality class. We found that the tape-peeling model belongs to the 1-dimensional DP universality class. 
Also, we reanalyzed the experimental results and found those results suggest that the tape-peeling trace is included in the 1-dimensional DP universality class. However the data for the 3 exponents are not found with a single experimental setting.
The detailed experiment for this finding is a future work.

\begin{table}[h]
  \caption{Scaling exponents.}
  \label{tbl:exponents}
  \begin{tabular}{cccc}
    \hline
     & $\beta$ & $\mu_\parallel$ & $\mu_\perp$\\
     \hline
  1-dimensional DP & $0.2765$ & $1.841$ & $1.748$  \\
  Tape-peeling model & $0.262(16)$ & $1.87(5)$ & $1.76(3)$\\
  Previous experiments & $0.275(29)$ & $1.80(7)$ & $1.72(6)$ \\
  \hline
  \end{tabular}
\end{table}

\clearpage

\section*{Acknowledgements}
We acknowledge JSPS KAKENHI 24K20863 for the financial support. We appreciate Prof. H. Nakao for the useful discussion. We also appreciate Prof. H. Kori and Prof. T. Yamaguchi for the discussion in the early stage of the research.

\bibliographystyle{junsrt}

\end{document}